\documentclass[twocolumn,showpacs,preprintnumbers,aps,nofootinbib]{revtex4-1}

\usepackage{graphicx}  
\usepackage{dcolumn}
\usepackage{bm}
\usepackage{amsmath,amssymb,amsfonts}
\usepackage{latexsym}
\usepackage{color}

\def\nn    {\nonumber}

\begin{document}

\title{The $tZH$ and $tZh$ production in 2HDM: Prospects for discovery at the LHC}
\author{Wei-Shu Hou and Tanmoy Modak}
\affiliation{Department of Physics, National Taiwan University, Taipei 10617, Taiwan}
\bigskip

\begin{abstract}
We study the discovery potential of the $cg \to tA\to tZH$ process at the LHC, 
where $A$ and $H$ are CP-odd and even exotic scalars, respectively.
The context is the general Two Higgs Doublet Model, where $cg \to tA$
is induced by the flavor changing neutral Higgs coupling $\rho_{tc}$.
We find that the process $cg \to tA\to tZH$ can be discovered for $m_A \sim 400$ GeV, 
but would likely require high luminosity running of the LHC. 
Such a discovery would shed light on the mechanism behind 
the observed Baryon Asymmetry of the Universe.
We also study $cg \to tA\to tZh$, where $h$ is the observed 125 GeV scalar,
but find it out of reach at the LHC. 

\end{abstract}

\maketitle

\section{Introduction}

The discovery of the Higgs boson~\cite{h125_discovery} $h(125)$ 
at the Large Hadron Collider (LHC) confirms the Standard Model (SM) 
as the correct theory at the electroweak scale.
As all fermions come in three copies, additional scalars might well exist in Nature. 
In particular, given that $h$ belongs to a weak doublet $\Phi$,
extra scalar doublets ought to be searched for.
However, the  apparent absence of New Physics (NP) so far at the 
LHC and the emergent ``approximate alignment'', i.e. 
the $h$ boson is found to resemble rather closely the SM Higgs boson, 
suggest that the extra scalars might be rather heavy. 
In this so-called decoupling limit~\cite{Gunion:2002zf}, 
where the exotic scalars are multi-TeV in mass, discovery becomes 
rather difficult even for the High Luminosity LHC (HL-LHC).

By adding just one scalar doublet $\Phi'$, 
the two Higgs doublet model (2HDM)~\cite{Branco:2011iw} 
is one of the simplest extensions of SM. 
We are interested in sub-TeV exotic scalars $A$, $H$, and $H^+$. 
The most popular 2HDMs, of interest already before the $h$ boson discovery,
are those with a $Z_2$ symmetry imposed~\cite{Branco:2011iw}. 
The $Z_2$ symmetry enforces the up- and down-type quarks to 
couple to just one scalar doublet,  thereby ensuring 
Natural Flavor Conservation (NFC)~\cite{Glashow:1976nt} 
and forbids all flavor changing neutral Higgs (FCNH) couplings. 
But this removes the possibility of {\it any} additional Yukawa coupling.

Our context is the general 2HDM (g2HDM), without imposing $Z_2$ symmetry. 
Indeed, approximate alignment can be accommodated~\cite{Hou:2017hiw,appalignment} 
without taking the decoupling limit, even with $\mathcal{O}(1)$ 
extra Higgs quartic couplings, clearing the way for sub-TeV $A$, $H$, and $H^+$. 
In the absence of $Z_2$ symmetry, both 
doublets couple to $u$- and $d$-type quarks,
and two separate Yukawa matrices $\lambda_{ij}^F =({\sqrt{2}m_i^F}/{v})\, \delta_{ij}$  
(with $v \simeq 246$ GeV) and $\rho_{ij}^F$ emerge after
diagonalization of the fermion mass matrices. Here, 
$F$ denotes $u$- and $d$-type quarks and $e$-type leptons,
with the fermion mass and mixing structure and approximate alignment 
together replacing the NFC condition~\cite{Hou:2017hiw}.
The $\lambda$ matrices are real and diagonal,
but the $\rho$ matrices are in general \textit{non-diagonal} and \textit{complex}.  
It was pointed out recently that $\mathcal{O}(1)$ $\rho_{tt}$ and $\rho_{tc}$
can drive electroweak baryogenesis (EWBG) rather  efficiently~\cite{Fuyuto:2017ewj,deVries:2017ncy}.

If $\rho_{tt}$ and $\rho_{tc}$ are $\mathcal{O}(1)$, 
one might discover the exotic scalars via the 
$cg\to t A/H \to t t \bar c$ process with clean same-sign top 
signature~\cite{Kohda:2017fkn,Hou:2018zmg}
 (see also Refs.~\cite{Hou:1997pm,Iguro:2017ysu,Altmannshofer:2016zrn}),
and also with $A/H\to t \bar t t$, i.e. the triple-top process~\cite{Kohda:2017fkn}. 
Induced by only $\rho_{tc}$, 
the same-sign top process might emerge already with full Run-2 data. 
On the other hand, the more exquisite triple-top process, 
which depends on both $\rho_{tt}$ and $\rho_{tc}$ couplings,
may require the inclusion of Run 3 data to show any indication. 
But if $\rho_{tt}$ is negligibly small, the triple-top discovery would not be possible.
In this paper we consider the case where 
$\rho_{tc}$ is $\mathcal{O}(1)$ but $\rho_{tt}$ is tiny, 
where another novel discovery mode would be $cg\to t A \to t ZH$
 (charge conjugate process always implied) for $m_A>  m_Z + m_ H$. 
With no dilution from $A\to t \bar t$, the process can provide 
an additional discovery mode that is complementary to 
Refs.~\cite{Kohda:2017fkn,Hou:2018zmg}, and
provide additional information on $\rho_{tc}$ driven EWBG.

The $cg\to t A \to t ZH$ process can be searched for 
in the inclusive $pp\to tA + X \to t Z H +X$ process, 
with $Z\to \ell^+\ell^-$, $H \to \bar t c + t \bar c$, 
and at least one top decaying semileptonically. 
We call this the $tZH$ process, 
the observation of which has another intriguing impact. 
It has been shown that the $A\to ZH$ decay can provide 
a smoking gun signature for the 
strongly first order electroweak phase transition (EWPT) which might have 
occurred in the early Universe~\cite{Turok:1990zg,Fromme:2006cm,Dorsch:2014qja}. 
A strongly first order EWPT is needed for the out of equilibrium condition 
that is required for successful EWBG~\cite{Cohen:1993nk}. 
Realizing the importance~\cite{gg2ZH}, indeed both ATLAS and CMS have 
pursued $gg\to A\to ZH$ search~\cite{Aaboud:2018eoy,Khachatryan:2016are}.
However, if $\rho_{tt}$ is tiny, $gg\to A$ vanishes,  
and the $tZH$ process will be a unique probe of the strongly first order EWPT mechanism, 
as well as the $\rho_{tc}$ driven EWBG scenario.

For completeness, we also study the prospect for 
the $cg\to t A \to t Zh$ process. 
The process is also induced by $\rho_{tc}$, but 
would depend on $\cos\gamma$, the $h$--$H$ mixing angle. 
The process can be searched for via $pp\to tA + X \to t Z h +X$,
with $t \to b \ell^+ \nu_\ell $,  $Z\to \ell^+\ell^-$ and $h\to b\bar b$, 
which we call the $tZh$ process.
It provides another complementary probe of the $\rho_{tc}$ driven EWBG scenario,
as well as the $c_\gamma$ mixing angle if $\rho_{tt}$ is rather small.

In the following, we first discuss the framework in Sec.~\ref{frame},
followed by the parameter space and discovery potential 
of the $tZH$ process in Sec.~\ref{proctZH}. 
Sec.~\ref{proctZh} is dedicated to the $tZh$ process, and 
we summarize our results with some discussion in Sec.~\ref{summ}.

\section{Framework}\label{frame}

The scalars $h$, $H$, $A$ and $H^+$ couple to fermions 
by~\cite{Davidson:2005cw}
\begin{align}
\mathcal{L} = 
&-\frac{1}{\sqrt{2}} \sum_{F = U, D, L'}
 \bar F_{i} \bigg[\big(-\lambda^F_{ij} s_\gamma + \rho^F_{ij} c_\gamma\big) h \nn\\
 &+\big(\lambda^F_{ij} c_\gamma + \rho^F_{ij} s_\gamma\big)H -i ~{\rm sgn}(Q_F) \rho^F_{ij} A\bigg]  R\; F_{j}\nn\\
 &-\bar{U}_i\left[(V\rho^D)_{ij} R-(\rho^{U\dagger}V)_{ij} L\right]D_j H^+ \nn\\
 &- \bar{\nu}_i\rho^L_{ij} R \; L'_j H^+ +{\rm H.c.},\label{eff}
\end{align}
where $L,R\equiv (1\mp\gamma_5)/2$, $i,j = 1, 2, 3$ are generation indices, $V$ is  Cabibbo-Kobayashi-Maskawa matrix,
$c_\gamma = \cos\gamma$ is the $h$--$H$ mixing angle between CP-even scalars, 
and $U=(u,c,t)$, $D = (d,s,b)$, $L'=(e,\mu,\tau)$ and $\nu=(\nu_e,\nu_\mu,\nu_\tau)$ are in vectors in flavor space. 
The matrices $\lambda^F_{ij}\; (=\sqrt{2}m_i^F/v)$ are real and diagonal,
whereas $\rho^F_{ij}$ are in general complex and non-diagonal.

In the Higgs basis, the most general $CP$-conserving two Higgs doublet potential 
can be written as~\cite{Davidson:2005cw, Hou:2017hiw}
\begin{align}
 & V(\Phi,\Phi') = \mu_{11}^2|\Phi|^2 + \mu_{22}^2|\Phi'|^2
            - (\mu_{12}^2\Phi^\dagger\Phi' + h.c.) \nn \\
 & \quad + \frac{\eta_1}{2}|\Phi|^4 + \frac{\eta_2}{2}|\Phi'|^4
           + \eta_3|\Phi|^2|\Phi'|^2  + \eta_4 |\Phi^\dagger\Phi'|^2 \nn \\
 & + \left[\frac{\eta_5}{2}(\Phi^\dagger\Phi')^2
     + \left(\eta_6 |\Phi|^2 + \eta_7|\Phi'|^2\right) \Phi^\dagger\Phi' + h.c.\right],
\label{pot}
\end{align}
where the vacuum expectation value $v$ arises from the doublet $\Phi$ 
via the minimization condition $\mu_{11}^2=-\frac{1}{2}\eta_1 v^2$, 
while $\left\langle \Phi'\right\rangle =0$ (hence $\mu_{22}^2 > 0$), 
and $\eta_i$s are quartic couplings. 
Here we follow the notation of Ref.~\cite{Hou:2017hiw}.
A second minimization condition, $\mu_{12}^2 = \frac{1}{2}\eta_6 v^2$, 
removes $\mu_{12}^2$, and 
the total number of parameters are reduced to nine~\cite{Hou:2017hiw}.

Two relations~\cite{Hou:2017hiw} arise for the mixing angle $\gamma$ 
when diagonalizing the mass-squared matrix for $h$, $H$,
%
\begin{align}
 c_\gamma^2 = \frac{\eta_1 v^2 - m_h^2}{m_H^2-m_h^2},~\quad \quad \sin{2\gamma} = \frac{2\eta_6 v^2}{m_H^2-m_h^2}.
\end{align}
The alignment limit, $c_\gamma \to 0$, is reached for 
$\eta_6 \to 0$~\cite{Hou:2017hiw}, hence $m_h^2 \to \eta_1 v^2$,
or via decoupling~\cite{Gunion:2002zf}, i.e. $m_H^2~\gg~v^2$.
But for small but not infinitesimal $c_\gamma$, 
one has $ c_\gamma \simeq|\eta_6| v^2/(m_H^2 -m_h^2)$.
This is the so-called approximate alignment~\cite{Hou:2017hiw}, 
i.e. small $c_\gamma$ values can be attained 
with $\eta_6,\, \eta_1 > m_h^2/v^2$.
The scalar masses can be expressed 
in terms of the parameters in Eq.~(\ref{pot}),
\begin{align}
 &m_{h,H}^2 = \frac{1}{2}\bigg[m_A^2 + (\eta_1 + \eta_5) v^2\nn\\
 &\quad\quad \quad\quad \mp \sqrt{\left(m_A^2+ (\eta_5 - \eta_1) v^2\right)^2 + 4 \eta_6^2 v^4}\bigg],\\
 &m_{A}^2 = \frac{1}{2}(\eta_3 + \eta_4 - \eta_5) v^2+ \mu_{22}^2,\\
 &m_{H^\pm}^2 = \frac{1}{2}\eta_3 v^2+ \mu_{22}^2.
\end{align}

The processes of interest are $cg\to tA \to t ZH$ and $t Zh$, 
where $cg\to tA$ is induced by $\rho_{tc}$, but
the $A \to Z H, Z h$ decays 
via the gauge couplings~\cite{Djouadi:2005gj,Branco:2011iw}
\begin{align}
 \frac{g_2 }{2 c_W}Z_\mu \left[c_\gamma (h \partial^\mu A  - A \partial^\mu h) 
 -s_\gamma(H \partial^\mu A - A \partial^\mu H)\right],
\label{zhlagra}
\end{align}
with $c_W$ the Weinberg angle and $g_2$ the $SU(2)_L$ gauge coupling.
We see from Eq.~\eqref{zhlagra} that 
$A \to ZH$ is proportional to $s_\gamma$, while 
$A\to Z h$ is proportional to $c_\gamma$. 
The coupling $\rho_{ct}$ can also generate $cg\to tA$, but 
it is very stringently constrained by flavor physics~\cite{Altunkaynak:2015twa}. 
We set $\rho_{ct}$ to zero throughout the paper for simplicity.

For nonzero $\rho_{tc}$, we remark that the discovery at LHC,
 if at all, would first occur through the
 $cg\to tA \to  t t \bar c$ process~\cite{Kohda:2017fkn, Hou:2018zmg}. 
For $m_A < 2 m_t$, if other $\rho_{ij}$s are small, 
$cg\to tA \to t ZH$ could be the only process to emerge after $cg\to tA \to  t t \bar c$. 
For $m_A > 2 m_t$, 
$cg\to tA \to  t t \bar c$ would in general be accompanied by the 
$cg\to tA \to  t t \bar t$ process~\cite{Kohda:2017fkn}, 
unless $\rho_{tt}$ is negligibly small, which we shall assume.
We shall focus on $t\to b \ell^+\nu_\ell$, $H\to t \bar c + \bar t c$, 
and $Z\to \ell^+\ell^-$ decays, with the top quark from $H$ decay 
also decaying semileptonically.
Thus, following a possible $cg\to tA \to  t t \bar c$ discovery,
$cg\to tA \to t ZH$ could be the only process that might provide
a complementary probe of the $\rho_{tc}$ driven EWBG, 
even for approximate alignment (i.e.  small $c_\gamma$)~\cite{approxalign}. 
In the following, we assume $\rho_{tc}$ is the only 
non-zero coupling and set all other couplings to zero.
Their impact, however, will be discussed later in the paper.

The prospect for $cg\to tA \to t Zh$ closely depends on the mixing angle $c_\gamma$, 
vanishing for $c_\gamma \to 0$. 
For large $\rho_{tt}$, $gg\to A \to Z h $~\cite{Aaboud17Sirunyan19}
probes $c_\gamma$. 
For negligibly small $\rho_{tt}$,
the process $cg\to tA \to t Zh$ can provide unique probe of $c_\gamma$. 
We shall focus on $t\to b \ell^+\nu_\ell$, $h\to b \bar b$ and $Z\to \ell^+\ell^-$.

\begin{figure*}[tb]
\center
\includegraphics[width=.4 \textwidth]{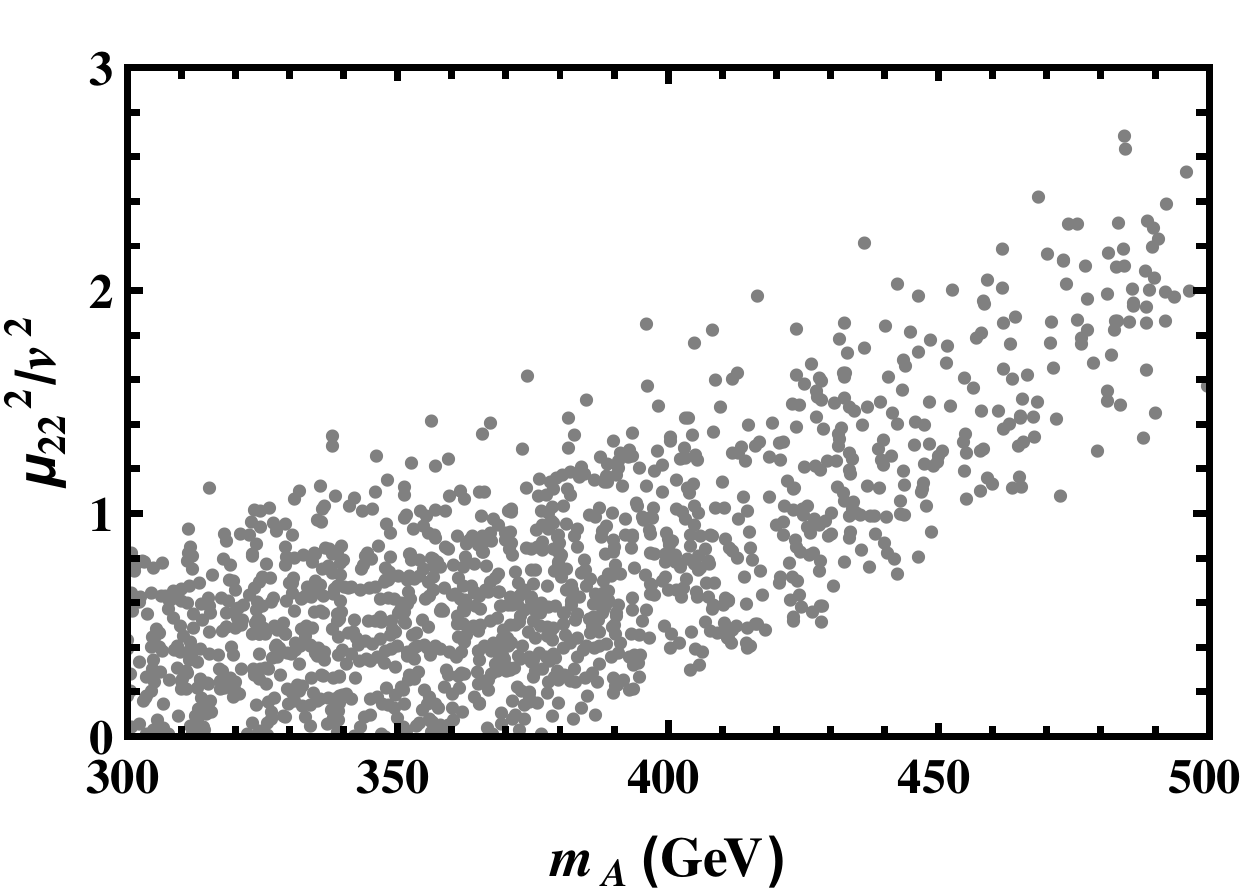}
\includegraphics[width=.4 \textwidth]{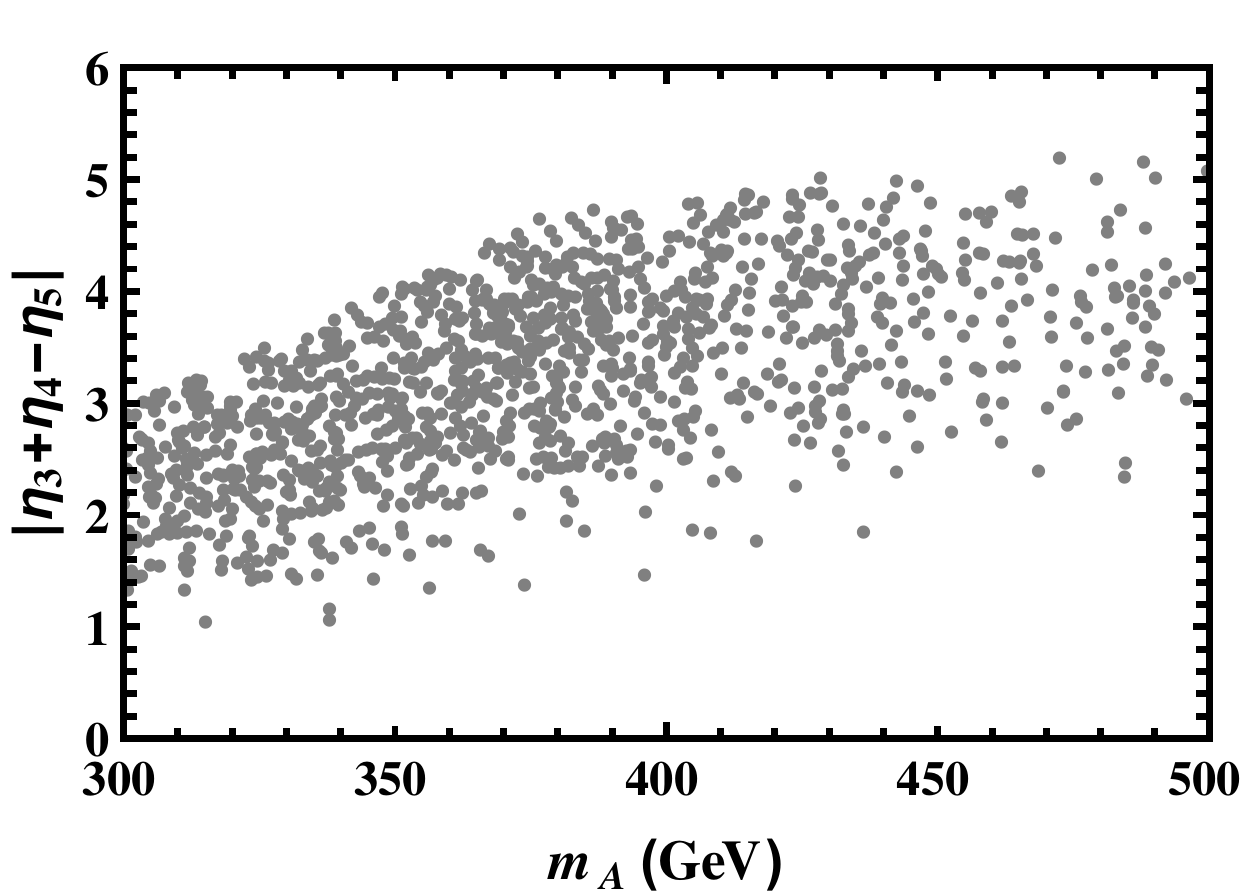}
\caption{The scanned points plotted in the $\mu_{22}^2/v^2$ vs $m_A$ (left) and $|\eta_3+\eta_4 -\eta_5|$ vs $m_A$  (right)  plane.} 
\label{scanplot_tZH}
\end{figure*}
\begin{table*}[tb]
\centering
\begin{tabular}{c |c| c| c| c | c | c| c | c |c| c|c| c c| cc}
\hline
BP  & $\eta_1$ &  $\eta_2$   &  $\eta_3$   & $\eta_4$  & $\eta_5$ &$\eta_{345}$ & $\eta_6$  & $\eta_7$  & $m_{H^\pm}$  & $m_A$ & $m_H$
&  $\frac{\mu_{22}^2}{v^2}$\\
 &&&&&&&&& (GeV) & (GeV) & (GeV)&\\ 
\hline
&&&&&&&&&&&&\\
$a$           & 0.258  & 2.133  & 2.87   & -0.569   & -1.194 & 1.107   & 0 & -0.791  & 310 & 339  & 207 & 0.15\\
$b$           & 0.258  & 1.366  & 2.718  & -0.733   & -1.97  & 0.015   & 0 & -0.252  & 354 & 404  & 208 & 0.71\\
$c$           & 0.258  & 2.432  & 2.67   &-0.652    & -2.21  & -0.192  & 0 &  0.091  & 393 & 449  & 260 & 1.21 \\
\hline
\hline
\end{tabular}
\caption{Parameter values for the three benchmark points. See text for details.}
\label{benchZH}
\end{table*}

\section{\boldmath 
The $tZH$ process}\label{proctZH}

In this section  we analyze the discovery potential of the $tZH$ process at the LHC. 
We first look at the relevant constraints on the parameter space, 
then find the discovery potential at $\sqrt{s}=14$ TeV. 
For simplicity, we assume all $\rho_{ij} = 0$ except $\rho_{tc}$.
However, the impact of other $\rho_{ij}$s will be discussed later in the paper. 
To simplify further, we set $c_\gamma = 0$ throughout this section.

\subsection{Parameter Space}

Let us find the available parameter space for the $tZH$ process. 
We first focus on the mass spectrum of the extra scalars $A$, $H$ and $H^+$. 
The process requires $A$ heavier than $H$ by at least $m_Z$. 
To find whether such mass spectrum exists, the dynamical parameters in Eq.~\eqref{pot} 
need to satisfy positivity, perturbativity, and tree-level unitarity conditions, 
for which we utilize 2HDMC~\cite{Eriksson:2009ws}. 
We first express the quartic couplings $\eta_1$, $\eta_{3{\rm -}6}$ in 
terms of~\cite{Davidson:2005cw,Hou:2017hiw} 
$\mu_{22}$, $m_h$, $m_H$, $m_A$, $m_{H^\pm}$, all normalized to $v$, 
and the mixing angle ${\gamma}$,
\begin{align}
& \eta_1 = \frac{m_h^2 s_\gamma^2 + m_H^2 c_\gamma^2}{v^2},\\
& \eta_3 =  \frac{2(m_{H^\pm}^2 - \mu_{22}^2)}{v^2},\\
& {\eta_4 = \frac{m_h^2 c_\gamma^2 + m_H^2 s_\gamma^2 -2 m_{H^\pm}^2+m_A^2}{v^2}},\\
& \eta_5 =  \frac{m_H^2 s_\gamma^2 + m_h^2 c_\gamma^2 - m_A^2}{v^2},\\
& \eta_6 =  \frac{(m_h^2 - m_H^2)(-s_\gamma)c_\gamma}{v^2}.
\end{align}
The quartic couplings $\eta_2$ and $\eta_7$ do not enter scalar masses, 
nor the mixing angle $\gamma$. 
Therefore in our analysis we take $v$, $m_h$, 
and $\gamma$, $m_A$, $m_H$, $m_{H^\pm}$, $\mu_{22}$, $\eta_2$, $\eta_7$ 
as the phenomenological parameters.

To save computation time, we randomly generate these parameters 
in the following ranges:
$\eta_2 \in [0, 3]$, $ \eta_7 \in [-3, 3]$, 
$\mu_{22} \in [0, 1000]$  GeV,
$m_A \in [300, 500]$ GeV,
$m_H \in [200, m_A - m_Z]$ GeV, 
$m_{H^\pm} \in [300, 500]$ GeV,
while satisfying $m_h = $  125 GeV.
Note that since the $cg\to tA \to tZH$ process depends only on $s_\gamma$,  
for simplicity we take $c_\gamma = 0$ in this section.
To simplify further, we demand $m_A < m_{H^\pm}+m_W$ 
to forbid the $A\to H^\pm W^\mp$ decay.
We then pass the randomly generated parameters to 2HDMC for scanning, 
which uses~\cite{Eriksson:2009ws} $m_{H^\pm}$ and $\Lambda_{1-7}$ 
as input parameters in the Higgs basis with $v$ as an implicit parameter. 
To match the 2HDMC convention, we identify 
$\eta_{1-7}$ as $\Lambda_{1-7}$ and take $-\pi/2\leq \gamma \leq \pi/2$,
and $\eta_2$ needs to be greater than zero as required by positivity, 
along with other more involved conditions in 2HDMC. 
In addition, we further conservatively demand all $|\eta_i| \leq 3$.

One also has to consider the stringent oblique $T$ parameter~\cite{Peskin:1991sw} 
constraint, which restricts the scalar masses $m_A$, $m_H$, and $m_{H^+}$~\cite{Froggatt:1991qw,Haber:2015pua}, 
and therefore the quartic couplings $\eta_i$s.  
We use the $T$ parameter expression given in Ref.~\cite{Haber:2015pua} 
and check that the points that passed positivity, unitarity and perturbativity 
conditions in 2HDMC, also satisfy the $T$ parameter constraint 
within $2\sigma$ error~\cite{Baak:2013ppa}.
These final points together are called ``scan points'', 
which are plotted as gray dots in Fig.~\ref{scanplot_tZH} 
in the $\mu_{22}^2/v^2$ and $|\eta_3+\eta_4 -\eta_5|$  vs $m_A$ planes.
The figure illustrates that there exists 
finite parameter space for $300~\mbox{GeV}\lesssim m_A \lesssim 500$ GeV.
which can facilitate $A\to Z  H$ decay. 
In general, heavier $m_A$ are possible, but the discovery potential 
diminishes with the rapid fall-off in parton luminosity.
From the scan points in Fig.~\ref{scanplot_tZH}, 
we choose three benchmark points (BPs) for our analysis, 
which are summarized in Table.~\ref{benchZH}. 

The coupling $\rho_{tc}$ is constrained by both LHC search and flavor physics.
As we assume $c_\gamma = 0$ throughout this section, the  most stringent limit 
arises from CMS search for four-top production~\cite{Sirunyan:2019wxt},
where the CRW region, i.e. Control Region for $t\bar tW$ background, 
gives the most relevant constraint. 
For non-zero $\rho_{tc}$, the process $cg\to t H/tA \to t t \bar c$ 
with same-sign top (same sign leptons plus jets) contributes abundantly 
to the CRW region, resulting in stringent constraint on $\rho_{tc}$.
There is, however, a subtlety. 
The $cg\to t H \to t t \bar c$ and $cg\to t A \to t t \bar c$ processes 
cancel each other exactly by destructive interference, if the masses and widths 
of $H$ and $A$ are the same~\cite{Kohda:2017fkn,Hou:2018zmg}.
This cancellation diminishes~\cite{Hou:2018zmg} when the 
$m_A-m_H$ mass splitting is larger than the respective widths, which is 
the case for all three BPs, where $m_A - m_H$ is more than 100 GeV. 
We refrain from a detailed discussion on the extraction procedure for this constraint, 
but refer the reader to Refs.~\cite{Hou:2018zmg,Hou:2019gpn}.
Following the procedure in Ref.~\cite{Hou:2018zmg} and utilizing the 
CRW region of Ref.~\cite{Sirunyan:2019wxt}, we find the 95\% CL upper limit on  
$\rho_{tc}$ are $0.4$, $0.5$, $0.45$ for the BP$a$,  BP$b$ and BP$c$ respectively. 

\begin{figure}[tb]
\center
\includegraphics[width=.4 \textwidth]{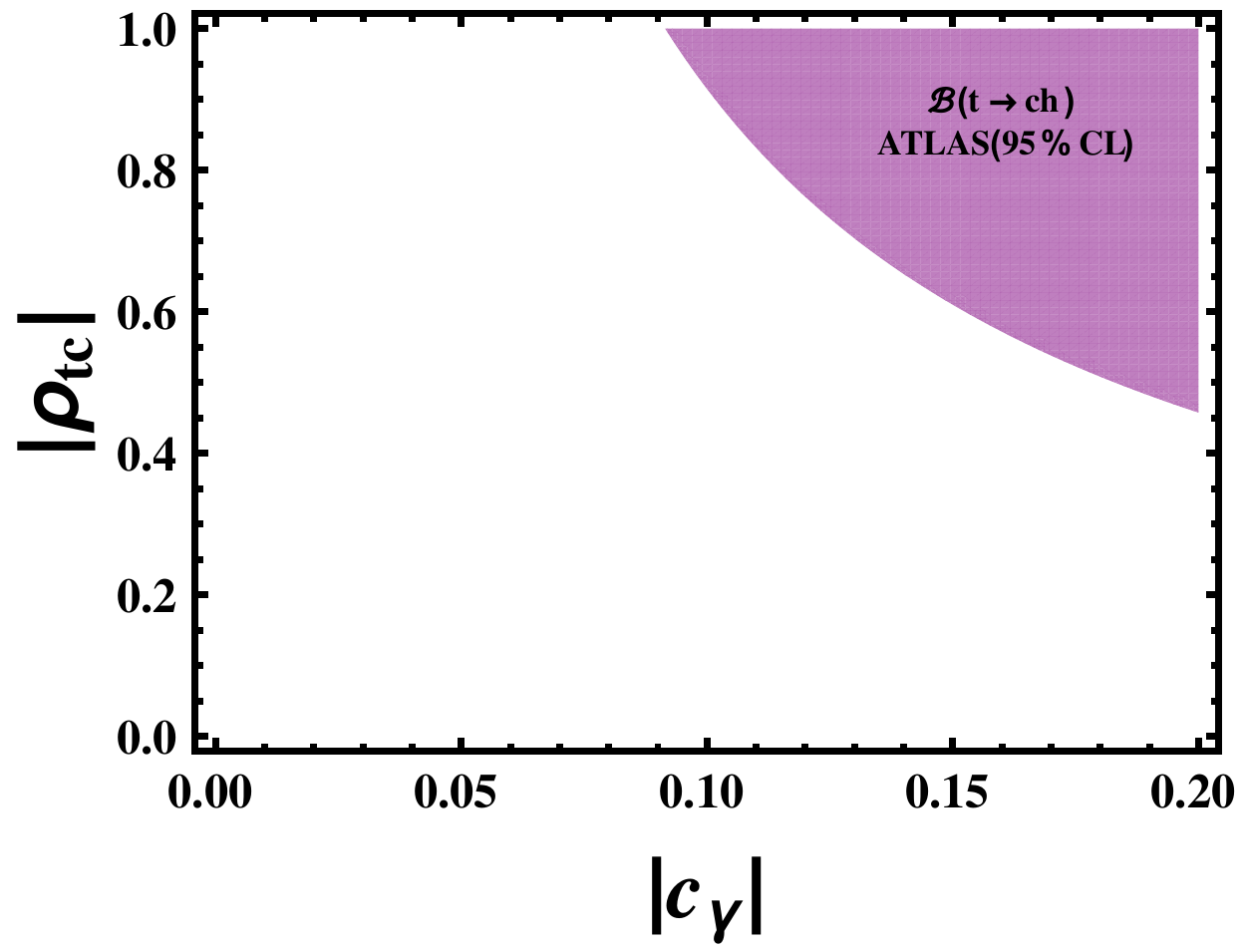}
\caption{
Constraint from $\mathcal{B}(t\to ch)$ measurement in $\rho_{tc}$ vs $c_\gamma$. } 
\label{ttoch}
\end{figure}

\begin{table}[t]
\centering
\begin{tabular}{|c| c| c |c |c  c }
\hline
  BP & \hspace{.1cm} $ \rho_{tc}$  \hspace{.1cm}
  &  \hspace{0.05cm} $\mathcal{B}(A \to t\bar c+\bar c t)$  \hspace{0.05cm}
  &  \hspace{.05cm}  $\mathcal{B}(A\to ZH)$ \hspace{.05cm} \\
\hline
\hline
     $a$           & 0.4           & 0.61                                & 0.39  \\
     $b$           & 0.5           & 0.41                                & 0.59     \\
     $c$           & 0.45          & 0.41                                & 0.59  \\            
\hline
\hline
\end{tabular}
\caption{Branching ratios for the benchmark points.} 
\label{branch}
\end{table}

The constraints from $\mathcal{B}(B\to X_s\gamma)$ and $B_q$ mixing ($q =d,s$) 
on $\rho_{tc}$ should also be considered, where $\rho_{tc}$ enters 
via $H^+$ coupling in the charm loop~\cite{Crivellin:2013wna,Li:2018aov}.
For example, reinterpreting the result of Ref.~\cite{Crivellin:2013wna}, one finds 
$|\rho_{tc}|\gtrsim 1$ is excluded for $m_{H^+}=300$ GeV from $B_s$ mixing, 
the ballpark mass range for $m_{H^+}$ for all three BPs. 
The constraints are weaker than those from the CRW region.
At this point we remark that, lighter $m_A$, $m_H$ and, $m_{H^\pm}$ compared to the three BPs
are also possible, but the constraints on $|\rho_{tc}|$ from CRW region, $\mathcal{B}(B\to X_s\gamma)$ and $B_q$ mixing
would be more severe.

For nonvanishing $c_\gamma$, $\rho_{tc}$ receives further constraints 
from $\mathcal{B}(t\to c h)$ measurement.
Although we set $c_\gamma = 0$ in this section, let us briefly discuss this constraint. 
Both ATLAS and CMS have searched for 
$t\to ch$ decay and set 95\% CL upper limits.
The latest ATLAS result is based on 36.1 fb$^{-1}$ data at 13 TeV, setting
the limit $\mathcal{B}(t\to c h) < 1.1\times 10^{-3}$~\cite{Aaboud:2018oqm}, 
while the CMS limit is $\mathcal{B}(t\to c h) < 4.7 \times 10^{-3}$~\cite{Sirunyan:2017uae}, based on 35.9 fb$^{-1}$. 
The ATLAS constraint on $\mathcal{B}(t\to c h)$~\cite{Aaboud:2018oqm} is 
illustrated in $\rho_{tc}$-$c_\gamma$ plane as the purple shaded region in Fig.~\ref{ttoch}, 
where we do not display the weaker CMS limit.
Taking $c_\gamma = 0.2$ for example, one gets
the upper limit of $|\rho_{tc}| \lesssim 0.5$ at $95\%$ CL~\cite{Hou:2019qqi},
but the limit weakens for smaller $c_\gamma$.

Under the assumptions made, there are only two decay modes, 
$A \to t\bar c + \bar t c$ and $A\to Z H$, for all three benchmark points. 
These branching ratios are summarized in Table~\ref{branch}, 
while $\mathcal{B}(H\to t\bar c + \bar t c)=1$. 
We note that for fixed $m_H$, $\mathcal{B}(A\to ZH)$ is larger for heavier $m_A$, 
hence $\mathcal{B}(A\to ZH)$ of BP$a$ is smaller than that of BP$b$.
The total decay widths of $A$ ($H$) for the three BPs respectively 
are 2.91 (0.18) GeV, 9.78 (0.29) GeV and, 9.65 (0.98) GeV.

\begin{figure*}[htb]
\center
\includegraphics[width=.47 \textwidth]{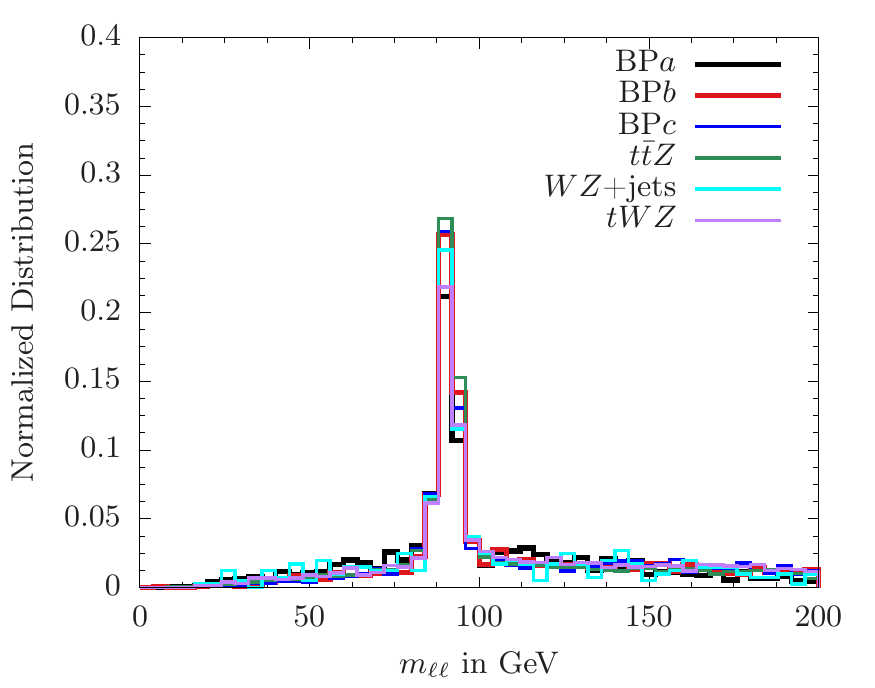}
\includegraphics[width=.47 \textwidth]{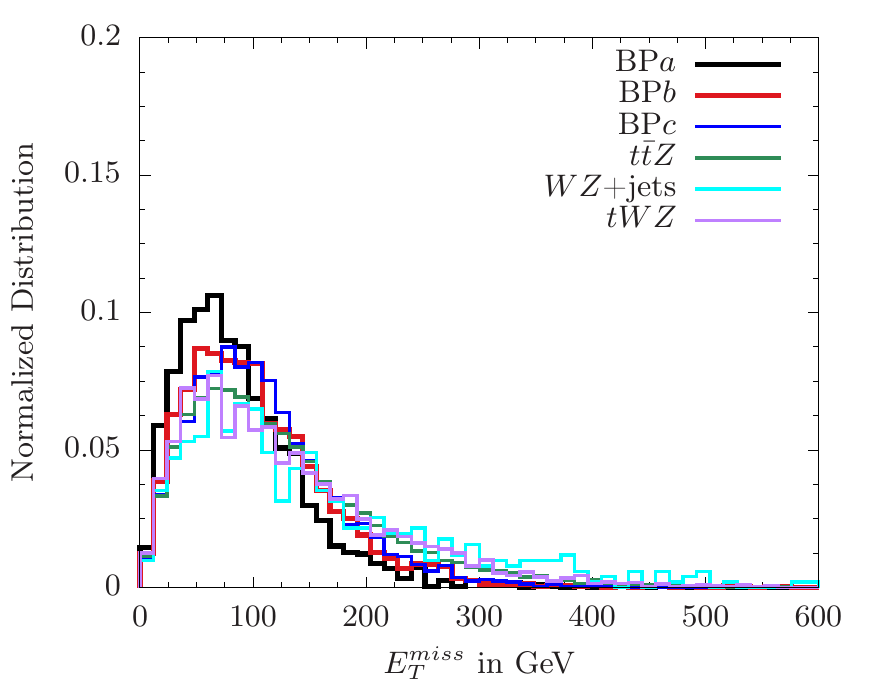}
\caption{The normalized $m_{\ell^+\ell^-}$ (left) and $E_T^{\rm miss}$ (right) distributions for the signal and background 
processes.} 
\label{mormdist}
\end{figure*}

\begin{table*}[hbt!]
\centering
\begin{tabular}{|c |c| c| c| c | c| c| c |c |c|}
\hline
&&&&&&&&&\\ 
      BP          & $t\bar tZ$  & $WZ+$      & $tWZ$      &  $4t$          &  $t\bar t h$    & $t t\bar W$ & $t Z$+          & Others & \ Total  \          \\
                  &             &jets        &            &                &                 &             & jets            &        &   Bkg.   \\      
\hline
\hline
&&&&&&&&&\\
       $a$         & 0.655        & 0.077      & 0.025     & 0.003          & 0.003           & 0.003        & 0.006          & 0.0001 &  0.772  \\ 
       $b$         & 0.902        & 0.11       & 0.035     & 0.004          & 0.004           & 0.004        & 0.007          & 0.0002 &  1.066   \\
       $c$         & 0.925        & 0.112      & 0.036     & 0.005          & 0.004           & 0.004        & 0.007          & 0.0002 &  1.093    \\ 
       
\hline
\hline
\end{tabular}
\caption{
Background cross sections (in fb) for the $tZH$ process 
after selection cuts at $\sqrt{s}=14$ TeV LHC.
The subdominant\ $3t+$jets, $3t+W$ are added together 
as ``Others'' in the second last column, 
while the last column is the total background. 
}
\label{bkgcross}
\end{table*}

\subsection{Collider Signature}

We now analyze the discovery prospects for $cg\to tA\to t ZH$ 
at the LHC with $\sqrt{s}=14$ TeV. 
The process can be searched for via
 $pp\to tA + X \to t Z H +X \to t Z (t \bar c + \bar t c) +X$,
with $Z\to \ell^+\ell^-$ and at least one of 
the final state top quarks decaying semileptonically.
$Z\to \tau^+ \tau^-$, $ \nu \bar \nu$ decays are also possible,
but we do not find them as promising.
The dominant backgrounds for the $tZH$ process 
arise from $t\bar tZ$ and $WZ+$jets processes, while 
$tWZ$, four-top quarks ($4t$), $t\bar t h$, $t t\bar W$ and $t Z$+jets are subdominant. 
Minor contributions come from  $3t+$jets and $3t+W$jets.

In order to find the discovery potential of the three benchmark points,
we generate background and signal event samples at LO
by Monte Carlo event generator MadGraph5\_aMC@NLO~\cite{Alwall:2014hca} 
with the parton distribution function (PDF) set NN23LO1~\cite{Ball:2013hta} at $\sqrt{s}=14$ TeV.
The event samples are then interfaced with 
PYTHIA~6.4~\cite{Sjostrand:2006za} for showering and hadronization, and 
finally fed into Delphes~3.4.0~\cite{deFavereau:2013fsa} 
to incorporate detector effects.
We have generated the matrix elements (ME) of signal and all backgrounds 
except for the $WZ+$jets with up to one additional jet in the final state, followed by ME and  
parton shower merging with the MLM matching scheme~\cite{Mangano:2006rw,Alwall:2007fs}. 
We considered two additional jets for ME and parton shower merging for $WZ+$jets
background.
We have not included backgrounds arising from the non-prompt and fake sources, as 
they are not properly modeled in Monte Carlo simulations, 
and usually require data to make estimates.
Here we have incorporated default ATLAS-based detector card available within Delphes framework. 
The effective model is implemented in FeynRules~\cite{Alloul:2013bka}.

The dominant $t\bar t Z$  cross section at LO is normalized to 
the NLO by the  $K$-factor 1.56~\cite{Campbell:2013yla}.
The $WZ+$jets background is adjusted to NNLO cross section 
by  a factor 2.07~\cite{Grazzini:2016swo}. Furthermore, 
the LO $\bar tZ +$ jets, $t\bar t h$, $4t$ and $t\bar{t} W^-$ ($t\bar{t} W^+$) 
cross sections are adjusted to NLO by $K$ factors 1.44~\cite{Alwall:2014hca}, 
1.27~\cite{twikittbarh}, 2.04~\cite{Alwall:2014hca} and 
1.35 (1.27)~\cite{Campbell:2012dh} respectively,
while the cross sections for $3t+$jets, $3t+W$jets and $tWZ$ are kept at LO.
For simplicity, the QCD correction factors for the  $tZj$ and $W^+Z+$jets processes are assumed
to be the same as their respective charge conjugate processes. The signal cross sections for all three BPs 
are kept at LO.

Let us discuss the event selection criteria for the $tZH$ process. 
Each event should contain at least three charged leptons  ($e$ and $\mu$),
at least three jets with at least two $b$-tagged, 
and missing transverse energy ($E_T^{\rm miss}$). 
The transverse momenta, $p_T$, of the leading charged lepton 
should be $> 25$ GeV, while the other two leptons should have $p_T > 20$ GeV.
The minimum transverse energy $E_T^{\rm miss}$ needs to be $> 35$ GeV.
All three jets are required to have $p_T > 20$ GeV. 
The absolute value of pseudo-rapidity, $|\eta|$, of the three leading leptons
and three jets (which includes two $b$-tagged jets) should be $< 2.5$. 
The separation $\Delta R$ between any two leptons, any two jets, 
and any jet and lepton should be $> 0.4$.
The jets are reconstructed by utilizing anti-$k_T$ algorithm 
with radius parameter $R = 0.6$.

The invariant mass of the two opposite-charge, same-flavor leptons, 
$m_{\ell^+\ell^-}$, is required to be within 
the $Z$ boson mass window $ 76 < m_{\ell^+\ell^-} < 100$ GeV. 
As there are at least three charged leptons in the event, 
with two coming from $Z$ decay and one from one of the $t$ quark decays, 
there will be at least two combinations of $m_{\ell^+\ell^-}$. 
We identify the pair having the invariant mass $m_{\ell^+\ell^-}$ 
closest to $m_{Z}$ as the one coming from $Z$ decay, 
and then impose the $m_{\ell^+\ell^-}$ mass cut. 
We finally veto events for $E_T^{\rm miss} > 150$ GeV, $250$ GeV and $270$ GeV 
for BP$a$, BP$b$ and BP$c$, respectively. 
The $E_T^{\rm miss}$ veto helps reduce the dominant $t\bar t Z$ background 
for all three BPs.

The normalized $m_{\ell^+\ell^-}$ and $E_T^{\rm miss}$ 
distributions before any selection cuts
 (with minimal default cuts during event generation in MadGraph5\_aMC@NLO)  
for the three BPs and backgrounds are plotted in Fig.~\ref{mormdist}. 

In this exploratory study, for simplicity we have not optimized the selection cuts 
such as $m_{\ell^+\ell^-}$ and $E_T^{\rm miss}$ for our BPs. 
The background cross sections after selection cuts
are summarized in Table~\ref{bkgcross} for all three BPs. 
In Table~\ref{signi} we give signal cross sections and 
the corresponding significance for the integrated luminosities
 $\mathcal{L}= 600$ and 3000 fb$^{-1}$.
The statistical significances in  Table~\ref{signi} are determined 
by using $\mathcal{Z} = \sqrt{2[ (S+B)\ln( 1+S/B )-S ]}$~\cite{Cowan:2010js},
where $S$ and $B$ are the number of signal and background events after selection.

\begin{table}[hbt!]
\centering
\begin{tabular}{|c |c| c| c | c }
\hline 
 &&\\
\hspace{.1cm}  BP   \hspace{.1cm}  \hspace{.1cm} & \hspace{.1cm} \ Signal \  \hspace{.1cm}& \hspace{.1cm} \ Significance ($\mathcal{Z}$) \hspace{.1cm}   \\ 
                                                 &    (fb)                                &  600 (3000) fb$^{-1}$    \\ 
 &&\\                                
\hline
\hline
                     $a$                       & 0.055                                  & 1.5  (3.4)    \\ 
                     $b$                       & 0.115                                  & 2.7  (6.0)    \\
                     $c$                       & 0.092                                   & 2.1  (4.8)     \\ 
\hline
\hline
\end{tabular}
\caption{
$tZH$ signal cross sections and significances after selection cuts 
for the three benchmark points.
}
\label{signi}
\end{table}

We find that the significances can reach up to $\sim1.5\sigma$, $2.7\sigma$ 
and $2.1\sigma$ for BP$a$, BP$b$ and BP$c$, respectively, for 600 fb$^{-1}$. 
With the full HL-LHC dataset (i.e. 3000 fb$^{-1}$ integrated luminosity) 
one can have $\sim3.4\sigma$, $6\sigma$ and $4.8\sigma$ for the BPs, respectively. 
With moderate $S/B\sim10\%$ for the three BPs, these significances illustrates that discovery 
is possible for $ m_A\sim 400$ GeV, while evidence is possible for $m_A\sim350$ GeV.
The significance is lower for lighter $m_A$ should not be surprising,
since $\mathcal{B}(A\to ZH)$ is lower for BP$a$ than BP$b$ and BP$c$. 
For heavier $m_A$ in BP$b$ and BP$c$, such enhancement in branching ratios 
can compensate lower $cg\to tA$ production cross section due to fall in parton luminosity.
Our results illustrate $\sim2\sigma$ hint is possible 
for $ m_A\sim 400$ GeV at Run 3 (300 fb$^{-1}$), 
but discovery would require the HL-LHC. 
The achievable significances depend mildly on the applied $E_T^{\rm miss}$ veto. 
E.g., if we apply the same $E_T^{\rm miss}$ veto that is chosen for BP$a$ to BP$b$ and BP$c$, the significances of the 
latter two BPs would drop by $\sim10\%$ and $\sim17\%$ respectively. However, rejecting events with  $E_T^{\rm miss} > 250$ GeV
would enhance the significance for BP$a$ by $\sim18\%$ but reduce by $\sim8\%$ for BP$c$, while keep the significance for BP$b$
unchanged. We remark in our exploratory analysis we have not optimized $E_T^{\rm miss}$ cut and leave out 
a more detailed analysis for future.

So far we have set all $\rho_{ij}=0$ except $\rho_{tc}$. Before closing 
this subsection, let us briefly discuss the impact of other  $\rho_{ij}$ couplings. 
If $\rho_{ij}$ follows similar flavor organization structure as in SM,
$\rho_{tt}$ could be $\mathcal{O}(\lambda_t)$, $\rho_{bb}\sim \lambda_b$, 
and $\rho_{\tau\tau}\sim \lambda_\tau$.
In general, presence of other $\rho_{ij}$s open up further decay modes of $A$ and $H$, 
which in turn dilutes $\mathcal{B}(A\to Z H)$, and hence 
the discovery potential of the $tZH$ process. 
For example, if $\rho_{tt} =\lambda_t$ (0.5), 
the achievable significances for BP$b$ and BP$c$ with full HL-LHC dataset
are reduced to $\sim2.7\sigma\, (4.6\sigma)$ and $1.7\sigma\, (3.3\sigma)$, 
respectively, due to non-zero $\mathcal{B}(A\to t\bar t)$. 
The significance of BP$a$ would remain unchanged as $m_A< 2m_t$.
Impact of other $\rho_{ij}$ couplings are significantly milder than $\rho_{tt}$. 
For example, for $\rho_{bb}\sim \lambda_b$ and $\rho_{\tau\tau}\sim \lambda_\tau$, 
the significance in Table~\ref{signi} remain practically the same.

Complex $\rho_{tt}$ provides a generally more robust mechanism for 
EWBG~\cite{Fuyuto:2017ewj,deVries:2017ncy}.
Having non-zero $\rho_{tt}$ motivates the conventional 
$gg\to H\to t \bar t$ scalar resonance search, 
or  $gg\to H t \bar t\to t \bar t t \bar t$~\cite{Craig} i.e. four-top search.
The former process suffers from large interference~\cite{Carena:2016npr} 
with the overwhelming $gg\to t\bar t$ background, 
leading to a peak-dip signature that makes detection difficult,
but recent searches by ATLAS~\cite{Aaboud:2017hnm} and 
CMS~\cite{Sirunyan:2019wph} find some sensitivity.
See Ref.~\cite{Hou:2019gpn} for a recent discussion in g2HDM context.
Presence of both $\rho_{tc}$ and $\rho_{tt}$ can induce 
$gg\to A/H\to t \bar c$~\cite{Altunkaynak:2015twa} 
and $cg\to t A / t H \to t t \bar t$  processes~\cite{Kohda:2017fkn} 
which can also be observable at the LHC,
but the former may suffer from $t+j$ mass resolution, 
which could be close to 200 GeV~\cite{KFC}.

\section{The $tZh$ process}\label{proctZh}

We now discuss the prospect of $tZh$ process, i.e. $pp\to tA + X \to t Z h +X$, with 
$t \to b \ell^+ \nu_\ell $,  $Z\to \ell^+\ell^-$ ($\ell = e,\mu$), and $h\to b\bar b$.
The process depends heavily on the mixing angle $c_\gamma$, as well as $\rho_{tc}$. 
In addition to the constraint from CMS CRW region~\cite{Sirunyan:2019wxt}, 
it also receives constraint from ATLAS 
$\mathcal{B}(t \to c h)$~\cite{Aaboud:2018oqm}. 
Indeed, larger $c_\gamma$ enhances $\mathcal{B}(A \to Z h)$, but 
$cg \to t A$ production is balanced by the stronger constraint on $\rho_{tc}$, 
as can be seen from Fig.~\ref{ttoch}. 
The process is further plagued by tiny $\mathcal{B}(Z\to \ell^+ \ell^-)$. 
These make the $tZh$ process not as promising as $tZH$ even for HL-LHC, 
which we make clear in the following.

To find the discovery potential, we choose a benchmark point 
where $A$ is heavier than $m_h+m_Z$, and lighter than $m_H+m_Z$. 
Such a choice would forbid $A\to Z H$ decay and enhance $\mathcal{B}(A \to Z h)$. 
Unlike the previous section, we also need $c_\gamma \neq 0$. 
We find such a benchmark point from 2HDMC which 
passes the perturbativity, unitarity, positivity constraints, 
as well as the $T$ parameter constraint. 
The parameter values are:
$\eta_1 = 0.428$, $\eta_2 = 2.88$, $\eta_3 = 0.795$, $\eta_4 = 2.916$, 
$\eta_5 = 2.334$,  $\eta_6= -0.897$, $\eta_7= 2.76$, 
$m_{H^+} = 378$ GeV, $m_A = 401$ GeV, $m_H = 559$ GeV,
 $c_\gamma = 0.186$ and  $\mu_{22}^2/v^2= 1.96$.
With this set of parameters, we find $\rho_{tc}$ values 
above $0.5$ is excluded at $95\%$ CL.
This is extracted from $\mathcal{B}(t \to c h)$~\cite{Aaboud:2018oqm}, while 
the constraint from CMS CRW region~\cite{Sirunyan:2019wxt} is a bit weaker. 
The branching ratios corresponding to this BP are 
$\mathcal{B}(A \to Z h)\approx 0.1$, $\mathcal{B}(A \to t \bar c + \bar t c)\approx 0.9$.

There exists several backgrounds for $tZh$ process. 
The dominant backgrounds are $t\bar t Z$, $4t$, $t \bar t h$, 
with subdominant backgrounds from  $tZ+$jets, $t\bar t W$,
$3t+$jets, $3t+W$jets and $tWZ$. 
To find the discovery potential, we follow the same procedure to 
generate signal and background events as in Sec.~\ref{proctZH}.
We keep signal cross section at LO, but for backgrounds 
we take the same QCD correction factors as in previous section.
The details of the selection cuts, and signal and background cross sections 
after selection cuts, are presented in an Appendix.

The statistical significance at $\sim 1.1\sigma$ turns out to be rather small, 
even with full HL-LHC dataset.
While significances would be lower for heavier $m_A$ 
due to fall in the parton luminosity, it does not improve much for lighter $m_A$. 
In the latter case, i.e. for lighter $m_A$, $\mathcal{B}(A \to Z h)$ becomes lower, 
and the constraint on $\rho_{tc}$ becomes more stringent from CMS CRW region~\cite{Sirunyan:2019wxt}.
For $cg\to t A\to t Zh$ search in $h \to W^+ W^{-*}$ and $Z\to b \bar b$ modes, 
one loses the mass reconstruction capability of $m_Z$, $m_h$ and $m_A$, 
hence the control of background processes.
Therefore, it is likely that the $tZh$ process woud 
remain below sensitivity even for HL-LHC.


\section{Discussion and Summary}\label{summ}

We have studied the discovery potential of $cg\to t A \to tZH$, $tZh$ processes at the LHC. 
The $tZH$ process can be discovered, albeit likely needing HL-LHC data.
Discovery is possible for $m_A\sim 400$ GeV, 
with statistical significance reaching up to $\sim6\sigma$ 
with full HL-LHC dataset. 
But $m_A$ cannot be much lighter or heavier than $\sim 400$ GeV. 
The discovery prospect for the $tZh$ process is rather limited, 
primarily due to the suppression from mixing angle $c_\gamma$
 (alignment ``protection''), 
and the constraint on $\rho_{tc}$ from $\mathcal{B}(t\to ch)$. 
With significance only about $1\sigma$ at best with 3000 fb$^{-1}$, 
$tZh$ seems out of reach at the LHC.
We note that the $cg\to t H \to tZA$ process is possible 
for $ m_H > m_Z + m_A$, and can be searched for by a strategy similar to $tZH$.
We also remark that $\rho_{tu}$ can induce $ug\to tA \to tZH$ process, 
with similar signature. 
Although $\rho_{tu}$ could become stringently constrained~\cite{Hou:2019uxa},  
the discovery potential is balanced by large valence-quark induced $ug\to tA$ production.

In general, the presence of $\rho_{tt}$ would 
reduce the discovery potential of $tZH$ because of $\mathcal{B}(A\to t \bar t)$, 
but it opens up other modes for $A\to ZH$ discovery, for example 
induce $A\to Z H$ signal via loop induced $gg\to A \to Z H$~\cite{gg2ZH}. 
The same is true for $\rho_{bb}$,
where $A\to ZH$ can be induced by $gb\to b A \to bZH$~\cite{Modak:2019nzl} as well 
as $gg\to b \bar b A \to  b \bar b ZH$~\cite{Modak:2019nzl,Modak:2018csw}.
One can also have $gg\to A \to Z h$~\cite{Aaboud17Sirunyan19} and 
$gg\to b \bar b A \to b\bar b Z h$~\cite{Aaboud17Sirunyan19,Ferreira:2017bnx,Coyle:2018ydo}. 
But both processes are again suppressed by the mixing angle $c_\gamma$. 
In general, the impact of $\rho_{bb}$ is inconsequential for the $tZH$ process, 
but the presence of $\rho_{tc}$ would reduce the discovery
potential for $\rho_{bb}$ induced $A\to ZH$ processes.

We have not discussed so far the uncertainties in our results.
We have not included QCD correction factors for signal in both the $tZH$ and $tZh$ processes.
In general, $c$-quark initiated processes have non-negligible  systematic uncertainties 
such as from PDF, which we have not included in our analysis.
Such uncertainties for $c$-quark initiated processes are discussed in Refs.~\cite{Buza:1996wv,Maltoni:2012pa}, 
while a detailed discussion of PDF choices and their uncertainties for Run 2 
can be found in Ref.~\cite{Butterworth:2015oua}. 
These lead to some uncertainties in our results.
A detailed estimate of such uncertainties is beyond the scope of this paper.

While the presence of $\rho_{tt}$ reduces the discovery potential 
of the $tZH$ process $m_A > 2 m_t$, it opens up the exquisite 
discovery mode $cg\to t A/ t H \to  t t \bar t$. 
It is also worthy of mention the ``excess'' 
seen by CMS~\cite{Sirunyan:2019wph} in $gg\to A \to t \bar t$ search at $m_A \approx 400$ GeV. 
Such excess can be interpreted within g2HDM framework~\cite{Hou:2019gpn},
if $\rho_{tt} \simeq 1.1$ and $\rho_{tc}\simeq 0.9$ 
with $m_{H^\pm} \gtrsim 530$ GeV and $m_H\gtrsim 500$ GeV.
Note that, for $\rho_{tt}\sim 1$, the $tZH$ discovery
 (or $cg\to t H \to t Z A$ discovery) is not possible 
due to suppression from $\mathcal{B}(A \to t\bar t)$ ($\mathcal{B}(H\to t \bar t)$) decay.
However, if this excess materializes into evidence or discovery by Run 3, 
$cg\to t A/ t H \to  t t \bar c$ might emerge immediately 
followed by discovery of $cg\to t A/ t H \to  t t \bar t$. 
%

In Summary, motivated by electroweak baryogenesis, 
we analyzed the discovery potential of the $cg\to t A \to t Z H$ process. 
Such process might be induced by extra Yukawa coupling $\rho_{tc} $
 if one removes the discrete $Z_2$ symmetry from 2HDM. 
We find discovery is possible at the HL-LHC if $m_A\sim 400$ GeV, 
but $\rho_{tt}$ would need to be small. 
For completeness, we have also studied the $cg\to t A \to t Z h$ process,
but do not find it promising. 
Discovery of the $cg\to t A \to t Z H$ process will not only 
shed light on the strongly first order electroweak phase transition,
it may also help uncover the mechanism behind 
the observed Baryon Asymmetry of the Universe. 

\vskip0.2cm
\begin{acknowledgments}
This research is supported by grants from MOST 106-2112-M-002-015-MY3,
107-2811-M-002-039, NTU 108L104019, 
and MOST 108-2811-M-002-537. 
\end{acknowledgments}

\appendix
\section{\boldmath Event selection for the $tZh$ process}\label{appen}

We discuss the event selection criteria and the corresponding 
signal and backgrounds for the $tZh$ process. 
Events are required to have at least three leptons, and
at least three $b$-jets with some missing transverse energy. 
The $p_T$ of the leading and other two subleading leptons are required to be $> 25$,
$20$ and $15$ GeV respectively, with pseudo-rapidity $|\eta| < 2.5$. 
The $p_T$  of all three $b$-jets are required to be $>20$ GeV with $|\eta| < 2.5$. 
The $E_T^{\rm miss}$ in each event should be  $> 35$ GeV.
We demand the separation $\Delta R$ between any 
two leptons, any two jets, and any jet and lepton to be $> 0.4$.
We then apply the $m_{\ell^+\ell^-}$ cut: for each event there are 
at least two possible $m_{\ell^+\ell^-}$ combinations, and
the $m_{\ell^+\ell^-}$ combination closest to $m_Z$ should be within $70~\mbox{GeV}<m_{\ell^+\ell^-}< 100$ GeV.
Similarly, there are at least two possible $m_{bb}$ combinations in each event.
We demanded the one that is closest to $m_h$ should be within $|m_{bb}-m_h|< 25$ GeV. 
Finally, we construct all possible $m_{\ell\ell bb}$ combinations 
from the three leading leptons and leading $b$-jets, 
and demand the $m_{\ell\ell bb}$ combination closest to $m_A$ should be within 
$|m_{\ell\ell bb}- m_A|<100$ GeV. 
The cross sections of signal and background processes 
after selection cuts are summarized in Table~\ref{sigbkgtZh}.

\begin{table}[h]
\centering
\begin{tabular}{|c |c| c| c| c | c| c| c |c |c|}
\hline
&&&&&\\ 
      Signal      & $t\bar tZ$      &  $4t$             & $t t\bar h$ & Others & \ Total  \          \\
       (fb)       &                 &                   &             &        &   Bkg.  \\                                    
\hline
\hline 
     0.003      & $\,0.025\,$      & $\,0.002\,$
      & $\,0.0001\,$     & $\,0.0001\,$  & 0.027 \\ 
\hline
\hline
\end{tabular}
\caption{Signal and background cross sections (in fb) for 
the $tZh$ process after selection cuts at $\sqrt{s}=14$ TeV LHC.
The subdominant backgrounds are added together as 
``Others'', and the last column is the total background. }
\label{sigbkgtZh}
\end{table}

 
\end{document}